\documentstyle[aps,prl,multicol,epsf]{revtex}
\begin{document}

\title{Access time of an adaptive random walk on the world-wide Web}

\author{Bosiljka Tadi\'c}

\address{Jo\v{z}ef Stefan Institute,
P.O. Box 3000, 1001 Ljubljana, Slovenia }

%\date\today

\maketitle
\begin{abstract}
We introduce and simulate the random walk that adapts move
strategies according to local node preferences on a directed
graph. We consider graphs with double-hierarchical connectivity
and variable wiring diagram in the universality class of the world-wide Web.
The ensemble of walkers reveals the structure of local subgraphs
with dominant promoters and attractors of links. The average access time
decays with the distance in hierarchy $\Delta q$ as a power
$<t_{aw}> \sim (\Delta q)^{-\theta }$. The access to highly
connected nodes is orders of magnitude shorter compared to the standard
random walk, suggesting the adaptive walk as an efficient
message-passing algorithm on this class of graphs.

 \end{abstract}

\begin{multicols}{2}
%\newline
%\section{Introduction}

Networks pervade all science \cite{Strogatz,Science} making the problem of
understanding  the structure and dynamics of complex networks
the greatest challenge today. Cellular and metabolic networks
\cite{cell}, chemical reactions \cite{Science,cell}, social collaboration
\cite{Mark} and  science citation networks \cite{SCN}, and the
world-wide Web \cite{WWW} are examples of networks that can be
characterized by random graphs with individual dynamics and coupling
architecture.  In his inspiring paper Strogatz \cite{Strogatz} suggests
that the inherent difficulty to understand networks lies in the
intimate relationship between their structural complexity and
evolution. To this end, the principle of {\it universality} that applies
to systems with scale-free structures can play an important role in
revealing those properties of the dynamics that are {\it relevant} for the
universal behavior of a network. The world-wide Web belongs to the class
of directed graphs with
double-hierarchical organization of node ranks \cite{WWW} in which the
wiring diagram rapidly changes in time. By exploiting the universality,
the recently proposed model \cite{BT} suggests minimum dynamic
rules that are able to account for the complexity of the Web. So far other
systems in this universality class are not known. It is tempting
to believe that certain catalytic reactions in open environment
\cite{DNA-motors} can be
represented by dynamic processes on directed graphs with variable wiring
diagram.

 Study of the response and relaxation of the network is the best way
 to understand how the structure  affects function. Various dynamic
processes, e.g., percolation and fragmentation
by diluting links \cite{percolation}(a), core percolation by stripping leaves
\cite{percolation}(b), and   spreading  of epidemics  \cite{Romu} on
networks of a given structure  were examined.

In this work we study the  random-walk dynamics on a directed graph. In
particular, we consider a double-hierarchical directed graph in the
universality class of the world-wide web \cite{BT}, in which in- and
out-links are governed by two distinct  (and statistically dependent)
power-law probability distributions.
We introduce a random walker which copies its
move strategies  from the local node linking preferences.
 An ensemble of adaptive random walkers  efficiently scans the
connected subgraphs.  The prominent feature of this class of
hierarchically   connected  graphs
is that the average access time to a node at distance $\Delta q $ measured
by in- or out-ranks decreases  as a  fractal power of the distance,
indicating the presence of a peculiar structure with few dominant promoter
and attractor nodes. The adaptive random walk
suggests how an efficient message-passing algorithm can be constructed,
that is driven by the  properties inherent to this class of graphs.
We demonstrate the advantages of the adaptive random walk by in parallel
simulating the standard random walk on the same graph.

We consider a directed graph evolving from the dynamic
rules which are recently suggested \cite{BT} to mimic growth of
the world-wide Web (the Web graph).
The basic properties of the model are \cite{BT}: (i) Directed  linking,
suggesting that  at a node out- and in-links are not symmetric;
(j) Growth and rearrangements (updates of links) at a unique time scale; and
(k) Bias update  and  bias attachment of links, with probabilities
specified below.
At each growth step a new node is added to the network and the number of
links changes by amount $M(t)$.
  A fraction $f_0(t)\equiv \alpha M(t)$ of new links are outgoing links
from the new added node $i=t$, whereas the remaining
$f_1(t)\equiv (1-\alpha )M(t)$ links are the updated links at other nodes
in the network. Hence, the relevant parameter in the model
 is the ratio of updated and added links at
each time step, i.e., $\beta \equiv f_1(t)/f_0(t) =(1-\alpha )/\alpha$,
which is independent of the actual increment $M(t)$.
Furthermore, the  variations in $M(t)$ are such that
 an average value $M\equiv {\overline{M(t)}}$ is finite,
which can be considered as a constant in first approximation. In practice,
the number of nodes {\it and} the number of links in the network increases
with time, so that reasonable values for the average $M$ are positive.
For consistency, we keep $M=1$ throughout this work (rendering reasonable
computation time) \cite{comment-M}.

Bias activity of agents who create or update  outgoing links from the Web
pages (nodes) and bias (preferential) attraction of links can be formulated
via following rules: At the growth step $i$ an outgoing link is created
from a node $n\leq i$ with  probability \cite{BT}
\begin{equation}
p_1(n,i) =     {{\alpha M + q_{out}(n,i)}\over{(1+\alpha)M*i}}  \ .
\label{out_linking}
\end{equation}
The link points towards the node $k$ with the probability
 \begin{equation}
p_2(k,i) =      {{\alpha M + q_{in}(k,i)}\over{(1+\alpha)M*i}} \ ,
\label{in_linking}
\end{equation}
where $q_{out}(n,i)$ and $q_{in}(k,i)$  are current  number of outgoing
and incoming links at respective nodes at the growth step $i$.  It is
assumed that at the time of addition of a node $i$ to the network
$q_{out}(i,i)=q_{in}(i,i)= 0$.   Therefore, the biasing in the dynamics
is linked to the time fluctuations of the node ranks. The effects of
the attachment rule in Eq.\ (\ref{in_linking}) to the distribution of
in-degree was studied analytically in Ref.\ \cite{DMS}.
For the values of the control parameter $\beta $ in the
range $0 <\beta < \infty $, corresponding to $1>\alpha > 0$, the network
has the capability to {\it rearrange } its structure of links at the pace
at which it grows. This  property makes the Web substantially different
from the networks that have static links. In our notation the networks
with fixed links, e.g., science citation network \cite{SCN},
correspond to the limit $\beta =0 $ (i.e., $\alpha = 1$).

As discussed in detail in Ref.\ \cite{BT} the network that evolves according
to the dynamic rules in Eqs.\ (\ref{out_linking})-(\ref{in_linking}) shows
a complex topology of links in which nodes are arranged hierarchically
both according to ranks of outgoing and incoming links, in
accordance with the data in the real Web \cite{WWW}. The cumulative
probability distributions that describe node ranks are \cite{PP-comment}
\begin{equation}
P(q_{out}) \sim q_{out}^{-(\tau _{out} -1)} \ ;\
P(q_{in}) \sim q_{in}^{-(\tau _{in}-1)} \ .
\label{PP}
\end{equation}
The corresponding scaling exponent $\tau _{in}$ is given by the exact result
\cite{DMS} $\tau _{in} = 2+\alpha $, whereas $\tau _{out} \approx 2+3\alpha $
\cite{BT} is approximately linear for $\alpha \leq 0.66$ and increases
faster than linear for $\alpha \to \alpha _c < 1$. For $\alpha \to 1$
the distribution of outgoing links loses the scaling behavior and approaches
random distribution. By comparison with the measured distributions
of outgoing and incoming links in the real world-wide Web \cite{WWW}
the parameter $\alpha $ is estimated as \cite{BT} $\alpha = 0.22 \pm 0.1$,
leading to $\beta $ in the range 3--4.

The topology of links in Eq.\ (\ref{PP}) affects the character of the
dynamic processes on the Web graph and  its relaxation properties. Next
we study two types of random walks on the Web graph---a naive random walk
and a walk with adaptive rules defined below.
We first grow a network of $N$ nodes using the
evolution rules in Eqs.\ (\ref{out_linking}-\ref{in_linking}).
The  walk then starts at time $t=0$ from a randomly selected initial node,
say node $n$. At this node we have $q_{out}(n) \equiv q_{out}(n,N)$ outgoing
links. (It is assumed that the network does not grow during
the walk, although this restriction is not essential for the results.)
A naive random walker selects its next move along one  of the outgoing
links of the node $n$ with  equal probability $w_0(n) = 1/q_{out}(n)$,
say the link pointing to the node $k$ and moves there. At the next step
 it makes a similar selection among  $q_{out}(k)$ links, and so on.
In contrast to this standard random walk rules, the  adaptive random walker
at each node selects the link with certain statistical
weight. Here we assume that the weights are correlated with the linking
preferences  in Eq.\ (\ref{in_linking}) of the visited node. In particular,
the walker investigates target nodes $k_\ell $ at the other end of each
the outgoing link $\ell $, $1\leq \ell \leq q_{out}(n)$ of the visited node
$n$ and assigns the weights $w_\ell $ to the corresponding links, where
\begin{equation}
w_\ell  \sim  p_2(k_\ell ,N) \ , ~~~\Sigma _{\ell =1}^{q_{out}(n)}
w_\ell =1 \ ,
\label{pw}
\end{equation}
and  $p_2(n,i=N)$ is given in Eq.\ (\ref{in_linking}) with the normalization
in Eq.\ (\ref{pw}). Thus, the adaptive walker uses the same principle of
selection that applied earlier to linking from the visited node.
 It should be noted that the weights $w_\ell $ are not
necessarily identical to linking probabilities, both because  they are
evaluated in fully grown  network and  normalized. It this way, the
adaptive random walker (ARW) utilizes the full information about
local architecture of in- and out-degree of the graph, whereas the naive
 random walker (NRW) is driven exclusively by the out-degree distribution,
thus exploiting only a part of the available information.
The walk continues as long as $q_{out}(k) > 0$ at last visited
node, and stops at a node with no out-links  $q_{out}(k) =0$
(border of the graph \cite{book}).

Both ARW and NRW on the graph traverse along
 a {\it connected path} of nodes, that, in principle, is  a subset of the
set of all connected nodes (so called connected component, which is usually
searched by the Web crawls). Similarly, the length of the walk is not
equal to the depth of the connected component, because the walk can move
backwards making  loops of any size. Hence the random-walk path
represents a local structure on the graph, that we discuss below.
In the metabolic and catalytic reaction networks the path of the walk
represents a possible  relaxation process between two
states corresponding to the departing and final node of the walk,
respectively. In this context a naive random walk can not be
considered as a process of choice, given that the presence of enzymes or
catalysts inevitably selects the preferred reaction, much similar to our
adaptive random walk.

It is  interesting to define the distance traversed by a walker
on the network in terms of the difference in node ranks $\Delta q$,
in which the graph has a nontrivial topology. The distribution of
such distances in principle depends on the time of the walk. In Fig.\ 1 we
present the results of time-integrated distribution $W(\Delta q)$ of
distances $\Delta q$ both for in- and out-links. Two bottom curves on the
main figure represent the distribution for the adaptive random walk, whereas
the curves above the dotted line are the corresponding distributions
in the case of the naive random walk. As it is seen immediately from Fig.\ 1
the connected subgraphs visited by an ensemble of random walkers
have topology that can be described by the power-law distributions
\begin{equation}
W(\Delta q _{out}) \sim (\Delta q _{out})^{-\delta _{out}} \ ;
W(\Delta q _{in}) \sim (\Delta q _{in})^{-\delta _{in}} \ ,
\label{Wq}
\end{equation}
with the distinct distribution of in- and out-degree, resembling
the global structure of the graph in Eq.\ (\ref{PP}). For the adaptive random
walk the exponents $\delta _{in}$ and $\delta _{out}$ are close to
$\tau _{in}$ and $\tau _{out}$ in Eq.\ (\ref{PP}),
respectively,  of the underlying graph structure (see inset to Fig.\ 1).
In fact, the scaling exponents do depend on the size of the ensemble $N_a$
relative to the size of the network $N$. In the inset to Fig.\ 1 we have
shown the exponents $\delta _{in}$ and $\delta _{out}$
measured by the ARW ensemble of the same size ($N_a=2\times 10^5$) as the
distributions in the main Fig.\ 1, but at larger network $N=10^5$. This
measurements result in larger exponents compared to the slopes measured
in smaller network  $N=10^4$ (cf. main Fig.\ 1). The exponents decrease
with increasing ratio $N_a/N$. When a large enough
ensemble of the adaptive random walkers is used the structure of
selected subgraphs approaches the underlying topology of the entire graph
 (see caption to Fig.\ 1). Therefore, an ensemble of
the adaptive random walkers on the Web graph can  be used as a  search
algorithm for the structure of connected subgraphs.
\narrowtext

\begin{figure}
\epsfxsize=82mm\epsffile[42 70 507 563]{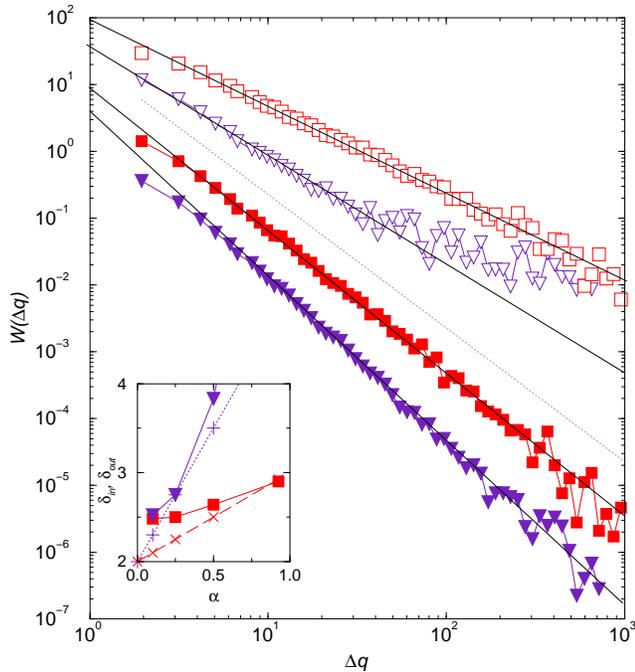}
%\vskip 8mm
\caption{\label{fig1}Time-integrated distributions $W(\Delta q)$ of
the distances $\Delta q$ measured in the node ranks plotted against
the distance $\Delta q$ for a naive random walk in-degree (open squares)
and out-degree (open triangles), and for adaptive random walk in-degree
(filled squares) and out-degree (filled triangles).  Data are log-binned
with bin ratio 1.1 .   Parameters: $\alpha =0.25$, $N=10^4$, $N_a =
2\times 10^5$. Slopes of the
curves are (top to bottom) $\delta _{in}^{nrw}=1.16$, $\delta _{out}^{nrw}=
1.55$, $\delta _{in}^{arw}=2.11$, $\delta _{out}^{arw}=2.62$.
Slope of dotted line is -2.
Inset: Scaling exponents of the ARW defined in Eq.\ (\ref{Wq}) for
out-degree (triangles) and in-degree (squares) for $N=10^5$,
$N_a = 2\times 10^5$ vs. $\alpha $. Also shown are the exponents $\tau _{in}$
(x) and $\tau _{out}$ (+) defined in Eq.\ (\ref{PP}).
 }
\end{figure}

In the case of naive random walk the distributions are qualitatively
different from the global graph  structure (see two top curves in main
Fig.\ 1). While the distributions of in- and out-degree still differ each
from other, the corresponding scaling exponents are $\delta _{in} <2$ and
$\delta _{out} <2$. (Some consequences of this property  will be discussed
later.)  The distributions $W(\Delta q)$ shown in Fig.\ 1
indicate that in the ensemble of  naive random walkers several well connected
nodes (i.e., for large $\Delta q$) are visited more frequently than in the
ensemble of the same size made of the adaptive random walkers (frequency can
differ up to four orders of magnitude for the simulated conditions,
see Fig.\ 1). This suggests that a naive random walker is wasting time
by walking in closed loops, which often pass through several highly
connected nodes on the graph.

Direct measurements of  the access time  support this conclusion.
We measure the average access time for a given distance $\Delta q$ in the
ensemble of walkers corresponding to the distributions in Fig.\ 1.
The results are shown in Fig.\ 2. The average
access time for the naive random walker is generally higher
than the one of the adaptive walker, the ratio
reaching  $<t _r>/<t_w> \sim 0.5\times 10^2  $ for large distances.
The most remarkable feature of this class of networks is that the average
access time {\it decreases} as a fractional power of the distance in
hierarchy, i.e.,
\begin{equation}
<t_w> \sim (\Delta q)^{-\theta }\Phi(t/\Delta q) \ ,
\label{tw}
\end{equation}
for distances $\Delta q$ below the  cut-off.  On the double-hierarchical
graphs that we study here, the Eq.\ (\ref{tw}) applies both for the random
and for the adaptive walk, with different exponents $\theta _{arw}$ and
$\theta _{nrw}$ as shown in Fig.\ 2.
This implies that decrease of the access time with the ranks differences
is an essential feature of these graphs that can be understood in
the following way. Consider a node of in-degree $k$. It can be directly
linked to a node of degree $k+Q$.  Majority of nodes have a link
with rather large rank difference $Q$---linking to a dominant
{\it attractor} in view of the rule (\ref{in_linking}).
The ARW, which is designed to follow such links, reaches quickly a
locally dominant attractor. According to the fast decaying distribution
$W(\Delta q)$ (we measured $\theta _{arw}^{in}=1.97\pm 0.04$) there is
a small number of attractors in the area scanned by the ARW. Hence, the
access to any other node, including a node with
a large out-degree often goes via a dominant attractor. The nodes with
a large out-degree have the capacity to disperse the links throughout
the network, because the probability to link back to the attractor decreases
with the number of out-links (cf. Eq.\ (\ref{pw}) ), i.e., they act as
the {\it promoters} of the dynamics.
In  Fig.\ 2 the average time to access a dominant promoter is 2-20 steps
for the ARW, compared with $\sim$ 200-600 steps for the NRW, suggesting that
the naive random walker makes up to 300 cycles containing
locally dominant attractor and promoter node.
Evidences of such structure in the real Web were recently discussed
in Ref.\ \cite{CLEVER}.

The probability  $P(t)$ that a walk survives for $t$ steps on the  Web
graph is a quantitative measure of  relaxation of the graph. The simulated
survival probability $P(t)$ shown in the
inset to Fig.\ 2 indicate once again
that the adaptive and naive random walk represent two types of
relaxation processes. Although the fitted expressions are
not definitive and require further theoretical analysis,
they suggest that a random walk on this class of graphs corresponds to a
stretch exponential relaxation, similar to relaxation in
complex disordered systems, whereas the adaptive random walk dies off
 nearly exponentially.
\begin{figure}
\epsfxsize=82mm\epsffile[42 70 507 563]{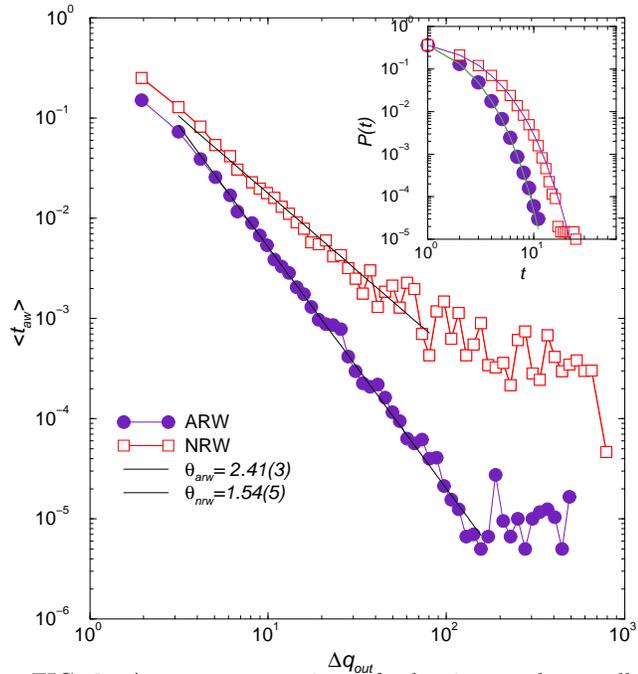}
%\vskip 8mm
\caption{\label{fig2}Average access time of adaptive random walker
(bullets) and a naive random walker (squares) versus distance $\Delta q_{out}$
measured by  out-degree ranks between departing and visited node. In both
cases an ensemble of $N_a=2\times 10^5$ walkers on the Web graph of $N=10^4$
nodes was used.
Inset: Survival probability $P(t)$ of time steps $t$ versus time steps
for adaptive (bullets) and naive (squares) random walker on the Web graph
simulated in the same conditions as the main figure. Fitting lines are:
(ARW) $P(t) = t^{-0.03}\exp{(-t^{0.99})}$, and (NRW)  $P(t) =
t^{+0.33}\exp{(-t^{0.82})}$.
 }
\end{figure}

As potential applications of these results we can mention the processes
of message-passing and infection-spreading on the Web graph. Assuming that
an infection can be transmitted with the walker, we find that both the complex
architecture of the graph and the  walk strategies are relevant for the
spreading. Due to a slow relaxation and heavy tail of the distribution
$W(\Delta q)$ of visited nodes ($\delta _{in} <2$) in the case of the naive
random walkers the epidemics is likely to spread over the entire graph.
With the adaptive  random walkers, on the other hand, the affected area
remains restricted, however, locally dominant nodes are  quickly affected.
The adaptive random walk also offers an efficient algorithm of message
passing to a given destination on the Web graph \cite{message-passing}.

To explore the complex structure of the Web graph we proposed an
adaptive random walk that learns its move strategies from the time varying
local dynamic rules of the graph itself. The walker has a short access
time to dominant nodes on the graph and affects a restricted area---the
properties that are relevant for the potential applications. The adaptive
random walk is a good candidate for a message-passing algorithm on the
Web graph (and catalytic reactions in the same universality class), which
builds its efficiency on fully exploiting the local graph structure
with double-hierarchical connectivity.

\acknowledgments
This work  was  supported by the Ministry
of Education, Science and Sports of the Republic of Slovenia. I thank
to Vyatcheslav Priezzhev for correspondence.

\end{multicols}

\begin{references}
\bibitem{comment-M}The actual number of links in the Web exceeds number
of nodes suggesting that a larger $M$ would be more realistic.
In the model the {\it universal} properties of the network in the scaling
region are not affected when $M$ is varied.
\bibitem{message-passing}Recently several other search algorithms that utilize
local information in a power-law graph  with symmetric links were proposed,
see   reference (c) in \cite{WWW}. It was shown that search time scales
sublinearly with the network size.
\end{references}
\end{document}